\documentclass[12pt,preprint]{aastex}
\usepackage{graphicx,epsf}
\shorttitle{Exposed Long-lifetime First-core}
\shortauthors{Tomida et al.}
\begin{document}
\title{Exposed Long-lifetime First-core: A New Model of First Cores Based on Radiation Hydrodynamics}

\author{Kengo Tomida\altaffilmark{1,2}, Masahiro N. Machida\altaffilmark{2}, Kazuya Saigo\altaffilmark{2}, Kohji Tomisaka\altaffilmark{1,2}, and Tomoaki Matsumoto\altaffilmark{3}}

\altaffiltext{1}{Department of Astronomical Science, The Graduate University for Advanced Studies (SOKENDAI), Osawa, Mitaka, Tokyo 181-8588, Japan; tomida@th.nao.ac.jp, tomisaka@th.nao.ac.jp}
\altaffiltext{2}{National Astronomical Observatory of Japan, Osawa, Mitaka, Tokyo 181-8588, Japan; masahiro.machida@nao.ac.jp, saigo.kazuya@nao.ac.jp}
\altaffiltext{3}{Faculty of Humanity and Environment, Hosei University, Fujimi, Chiyodaku, Tokyo 102-8160, Japan; matsu@hosei.ac.jp}

\begin{abstract}
A first adiabatic core is a transient object formed in the early phase of star formation. The observation of a first core is believed to be difficult because of its short lifetime and low luminosity. On the basis of radiation hydrodynamic simulations, we propose a novel theoretical model of first cores, Exposed Long-lifetime First core (ELF). In the very low-mass molecular core, the first core evolves slowly and lives longer than 10,000 years because the accretion rate is considerably low. The evolution of ELFs is different from that of ordinary first cores because radiation cooling has a significant effect there. We also carry out radiation transfer calculation of dust-continuum emission from ELFs to predict their observational properties. ELFs have slightly fainter but similar SEDs to ordinary first cores in radio wavelengths, therefore they can be observed. Although the probabilities that such low mass cores become gravitationally unstable and start to collapse are low, we still can expect that a considerable number of ELFs can be formed because there are many low-mass molecular cloud cores in star-forming regions that can be progenitors of ELFs.

\end{abstract}

\keywords{stars: formation --- ISM: clouds --- radiative transfer --- hydrodynamics}

\section{Introduction}

\citet{lrs69} studied star-formation processes in low-mass molecular cloud cores using one-dimensional (1D) spherically symmetric hydrodynamic simulations. In his work, a transient, quasi-hydrostatic object is formed in the early phase of the proto-stellar collapse, a so-called first (adiabatic) core. The first core evolves via accretion from the natal core and it collapses again when it attains a temperature high enough to dissociate hydrogen molecules. Although first cores are transient, they are essential to understand star formation processes because they assume a crucial role in the two major problems in the star formation process, the angular momentum and the magnetic flux problem. The first cores are interesting sites of diverse phenomena related to these problems such as driving molecular outflow \citep{tmsk02,mt04,mim06,dp09,tomida10a,cp10}, binary formation \citep{mtmi08,com10}, planet formation \citep{imm10}, circumstellar disk formation \citep{hc09,mim10,mt10,bate10,db10}, and so on. More than 40 years have passed since Larson's theoretical prediction, but it has not been robustly confirmed yet by observations due to its observational difficulties. The lifetime of a first core is so short, on the order of 100--1000 yrs, compared to the dynamical time-scale of its natal cloud, which can be about $10^4-10^6$ yrs depending on its initial density, that the detection probability of the first core is considered to be small. The first core is also deeply obscured in the infalling envelope with very large column density; hence it is believed to be difficult to observe with current instruments. However, if we could observe first cores directly, they will significantly contribute to our understanding of star formation processes.

The evolution of cloud cores with masses around $1M_\odot$ has been well studied using three-dimensional (3D) numerical simulations \citep{bate98,wb06,saigo08,com10,tomida10a,bate10}. In those cases, the first cores evolve over very short dynamical time-scales because accretion from their natal cores controls the evolution. Rotational support extends the lifetime of the first cores, but it is still brief compared to the dynamical time-scales of the clouds \citep{saigo08}. However, we can expect that the first cores formed in very low-mass cloud cores do not evolve into the second collapse stage over these short dynamical time-scales because most of the gas in the envelope quickly falls into the first core and the accretion rate rapidly decreases. These first cores survive longer before the second collapse starts than do cores formed in more massive clouds. In the long-term evolution of these systems, radiation cooling plays an important role in disk stability and angular momentum transport \citep{gm01}. To handle the gas thermodynamics properly, radiation hydrodynamics (RHD) simulations are required.

Very recently, several groups reported possible detection of first core candidates. \citet{chen10a} found a very faint and compact core and claimed that it was a good first core candidate. Their object must be very young, but it seems to be more evolved than a first core because it is associated with a well-collimated and high-velocity outflow, which is thought to be driven from a protostar after the second collapse. \citet{chen10b} reported another candidates in the R Corona Australis cloud. \citet{en10} found a good candidate for a first core whose spectral energy distribution (SED) is consistent with a theoretical model. A Japanese group (Kawabe et al., in preparation) also reported a peculiar dense core which is a good first core candidate. These objects need to be confirmed by more precise observations, but these observations imply that first cores are more common than expected. The observed candidates commonly have faint SEDs and some of them seem to have small masses. Considering observed core mass function (CMF), there are so many low-mass cores whose masses are less than $1M_\odot$ \citep{motte,en07,rath09,gould}. In the initial mass function (IMF), there are still many low-mass stars and brown dwarfs \citep{ch03}, although the relationship between the two mass functions is still unclear. Therefore it seems reasonable and interesting to study the evolution of low-mass cores which have masses less than $1M_\odot$.

The observational properties of first cores such as their SEDs were predicted on the basis of 1D RHD simulations \citep{mi00} and post-processing radiation transfer calculations by using the results of 3D hydrodynamic simulations with barotropic approximation \citep{st10}. First cores have faint, low-temperature black-body-like SEDs in wide wavelengths from mid-infrared to centi-meter-length radio wave. First cores are faint but detectable with modern facilities such as the {\it Herschel} Space Observatory, the Submillimeter Array (SMA), the Expanded Very Large Array and the Atacama Large Millimetre/sub-millimetre Array (ALMA). With ALMA in full operation, it is expected that we will be able to resolve the structure of first cores such as spiral-arms formed via gravitational instability in the disk and driving region of the molecular outflow driven by magnetic fields predicted with magnetohydrodynamics (MHD) simulations.

In this paper, we propose a new scenario for the evolution of first cores formed in very low-mass cloud cores. We prove that first cores in the very low mass molecular cloud cores have significantly long lifetimes. We also calculated SEDs and visibility amplitude distributions of our model, and compared them with the models of both a typical first core and a starless core. The plan of this paper is as follows. In section 2 we briefly describe our method of RHD simulations and models used in the simulations. We show the results of the numerical simulations in section 3 and the observational properties of those simulated first cores in section 4. Section 5 is devoted to conclusions and discussions.

\section{Method and Models}

We perform 3D nested-grid self-gravitational RHD simulations. Radiation transfer is treated with gray flux limited diffusion (FLD) approximation \citep{lp81}. The details of our simulation code are described in our previous work \citep{tomida10a}, but we neglect magnetic fields in this work because initially we just intend to prove the concept. We use an idealized equation-of-state (EOS) with adiabatic index $\gamma = 5/3$ which is different from that used in our previous work, $\gamma = 7/5$, to make the calculations easier. % This affects the results quantitatively, but the qualitative scenario is still valid.
 We use the compiled tables of Rosseland and Planck mean opacities of \citet{semenov} and \citet{fer05}, as in our previous paper. The number of grid points in each level of the nested-grids is $64^3$, and the finer grid is generated around the center of the computational domain to resolve the local Jeans length with at least 16 grid cells to prevent artificial fragmentations \citep{trlv97}.% Our code is optimized and parallelized for NEC SX-9 vector supercomputer, and the longest calculation for the low-mass core model takes more than 10,000 CPU hours because it is necessary to follow the long-term evolution of the first core. The finest resolution around the center of the cloud is $\sim 0.02 \  {\rm AU}$. 

We calculated three models for comparison: {\it S1} is a non-rotating $1M_\odot$ model, {\it R1} is a rotating $1M_\odot$ model, and {\it R01} is a rotating very low-mass model whose mass is $0.1M_\odot$. As the initial conditions, we take critical Bonnor-Ebert-like spheres with uniform rotation, and increase the density by a factor of 1.6 to make them unstable. The initial gas densities at the center and the radii of the clouds are $(\rho_c, R_c)=(3.2\times 10^{-18} \  {\rm g \,  cm^{-3}}, \  6300 \  {\rm AU})$ in {\it S1} and {\it R1}, and $(\rho_c, R_c)=(3.2\times 10^{-16} \  {\rm g \,  cm^{-3}}, \  630 \  {\rm AU})$ in {\it R01}. It is not trivial how we should scale the angular velocity between models with different masses, but here, we assume that both {\it R1} and {\it R01} initially have the same amount of rotational energy, $T \equiv \alpha MR^2\Omega^2$ ($\alpha$ is a constant of order unity). The initial angular velocities are $4.3\times 10^{-14} \  {\rm s^{-1}}$ and $1.4\times 10^{-12} \  {\rm s^{-1}}$ in {\it R1} and {\it R01}, respectively. We adopt the boundary conditions that all the cells outside the sphere of the cloud radius maintain their initial values. The initial gas temperature and boundary conditions for radiation transfer are set to 10K. Here, we note that our boundary conditions allow the gas to inflow into the computational domain through the boundaries, and at the end of the simulations about 30\% of the initial mass increased in the low-mass case and less than 10\% increased in the $1M_\odot$ cases.
%the two initial conditions are connected by homologous collapse and specific angular momentum conservation; $M\propto \rho r^3={\rm const}$ and $j\propto r^2\Omega$, then $\Omega\propto \rho^{-\frac{2}{3}}$.

\section{Results}

\begin{figure}[p]
\scalebox{0.2}{\includegraphics{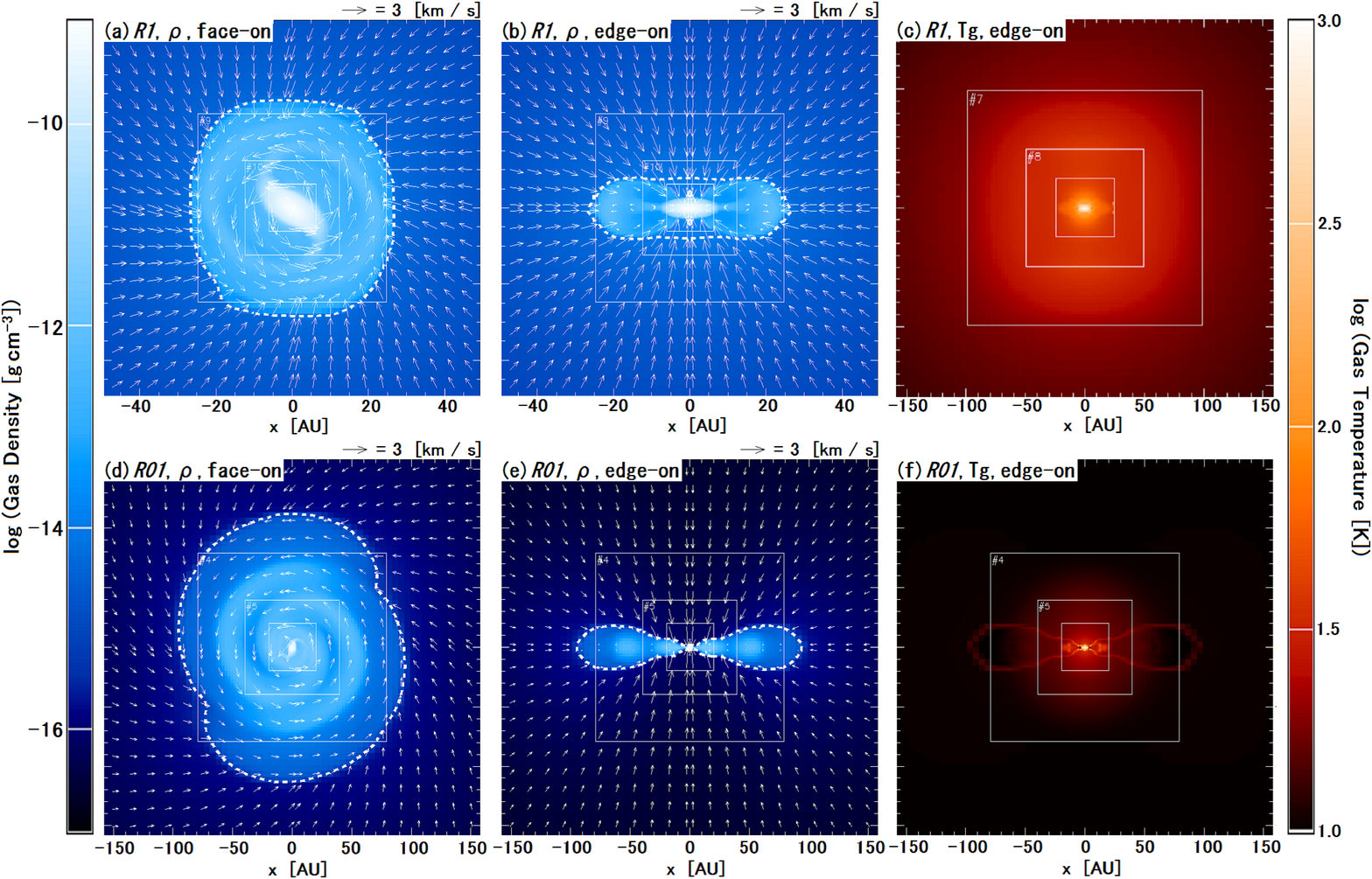}}
\caption{Horizontal (a, d) and vertical (b, c, e, f) cross sections of gas density (a, b, d, e) and temperature (c, f) at the epoch of the same first-core masses. (a)-(c): {\it R1} at $t_{\rm FC} = 3,100 \  {\rm yrs}$, (d)-(f): {\it R01} at $t_{\rm FC} = 10,600 \  {\rm yrs}$. The first cores are indicated by white dashed lines. $\rho_{\rm FC} = 1.0\times 10^{-14} \  {\rm g \  cm^{-3}}$ and $7.6\times 10^{-16} \  {\rm g \  cm^{-3}}$ in {\it R1} and {\it R01}, respectively. Despite the same first core masses, the envelope in {\it R01} is clearly depleted and colder. Note that the spatial scales in (a) and (b) are different from others. }
\label{fig1}
\end{figure}

We show the cross sections of the gas density and temperature of Model {\it R1} and Model {\it R01} at the epoch when they have the same first core masses, $1.06\times 10^{-1}M_\odot$, in Figure~\ref{fig1}. Here, the first core mass $M_{\rm FC}$ is defined as
\begin{equation}
M_{\rm FC} = \sum_{\rho>\rho_{\rm FC}}\rho \Delta V_c, 
\end{equation}
where $\Delta V_c$ is the volume of the cell and $\rho_{\rm FC}$ is the critical density defined as the minimum density in the gas that experienced the shock, which we identified as the region where the radial velocity is smaller than the local sonic speed, $|v_r| < c_s$. The age of the first core at this epoch, $t_{\rm FC}$, is 3,100 yrs in {\it R1} and 10,600 yrs in {\it R01} from the core-formation. Model {\it R01} has a significantly larger ($\sim 100$ AU) first core disk because it is more evolved and the gas with a larger specific angular momentum from the outer region has accreted onto the core. Another outstanding difference between the models can be seen in the gas density and temperature in their envelopes. For the low mass core model, almost all the materials in the natal core have already accreted onto the first core, and therefore, the gas density in the envelope is reduced drastically. Two factors cause the lower temperature in the envelope of {\it R01}: the smaller accretion rate in the very low-mass core results in the weaker shock at the surface of the first core, and thus, less entropy is produced at the shock (in other words, the first core is intrinsically colder and fainter), and the smaller optical depth of the envelope contributes to the efficient radiation transport and cooling. %These characteristics result in the differences in the SEDs between the two models.

\begin{figure}[p]
\scalebox{1}{\includegraphics{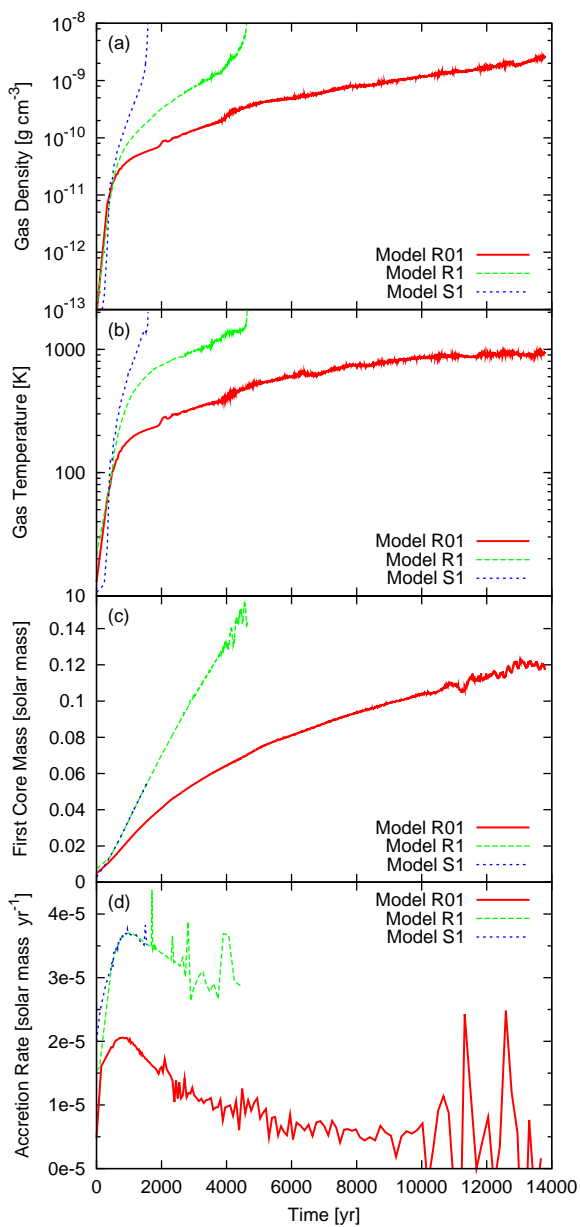}}
\caption{Time-evolution of the physical quantities, central gas density (a), temperature (b), first core mass (c), and smoothed accretion rate (d). Model {\it R01} shows prominently longer lifetime than $1 M_\odot$ models, {\it S1} and {\it R1}.}
\label{time}
\end{figure}

In Figure~\ref{time}, we show the time-evolution of physical quantities in each model such as central gas density (a), temperature (b), first core mass (c) and accretion rate onto the first core (d). It is obvious that {\it R01} has a smaller accretion rate and its mass increases more slowly than it does in other models. As a result, the gas density and temperature at the center of the first core evolve significantly slower. The lifetime of the first core, from its formation to the second collapse, reaches more than $10^4$ yrs in {\it R01}. Both $1M_\odot$ models achieved similar accretion rates because the initial structure of the cloud core determines the accretion rate \citep{saigo08}, although the non-rotating model evolves faster than the rotating model. These results clearly indicate that centrifugal force supports a considerable mass and prevents the first core from collapsing. We note that the lifetime of the non-rotating model is longer than previous predictions \citep{mi00}, by about a factor of two, because we assume a simple stiff EOS of $\gamma = 5/3$, and our model has a smaller accretion rate due to stable initial conditions. But in the rotating models, the difference of EOS affects the results less significantly because the centrifugal force dominates the structure of the first core disk.

\begin{figure}[p]
\scalebox{1}{\includegraphics{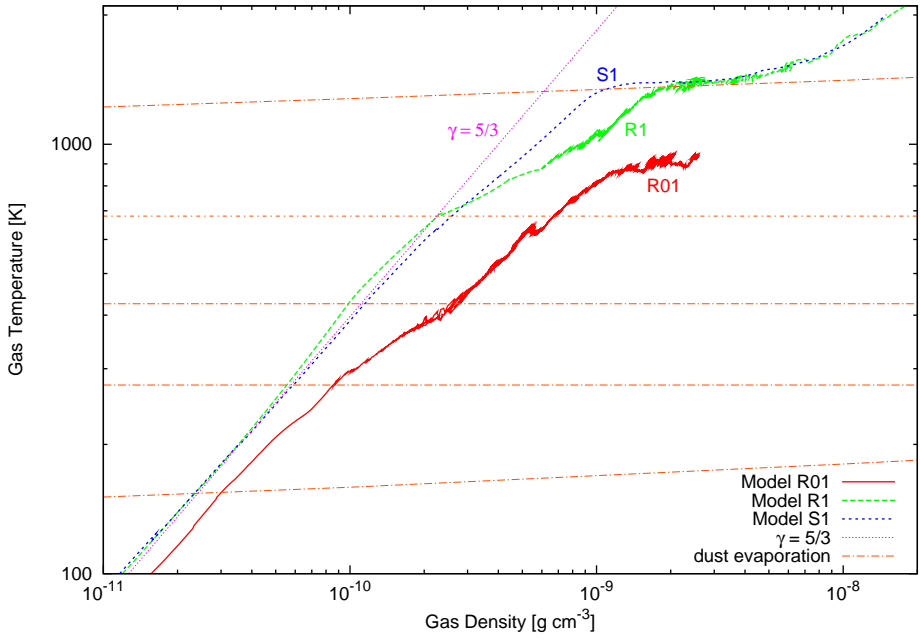}}
\caption{The evolution tracks of the thermal properties of the central gas elements in the $\rho-T$ plane. The dust evaporation temperatures are over-plotted with orange dash-dotted lines. All the models show the influence of radiation cooling but it is most significant in the low mass model.}
\label{rhot}
\end{figure}

We show the thermal evolution tracks of the gas elements at the center of the clouds in the $\rho-T$ plane in Figure~\ref{rhot}. All the models show the effects of radiation cooling, but it appears most significantly in the low mass model. We also plot the evaporation temperatures of each dust component, which affect the thermal evolution of the gas. When the gas temperature reaches the evaporation temperature of iron and silicates, $T\sim 1400 \  {\rm K}$, the opacity drops substantially and the core collapses violently, similar to the second collapse due to the endothermic reaction of hydrogen molecule dissociation. However, if we adopt a soft EOS with $\gamma = 7/5$, the impact of dust evaporation becomes less important because the gas density (and therefore the optical depth) at the same temperature is higher. Thus the effect of radiation cooling is important in the low mass core where dynamical accretion does not dominate the evolution. RHD simulations are required for these systems because the barotropic approximation fails to reproduce the realistic thermal evolution.

Our results show that the low-mass model has a longer lifetime compared to the dynamical time-scale of its natal core. This suggests that a large fraction of low-mass cores can harbor the first core when we carry out an unbiased survey of collapsing starless cores. 
Although only a small fraction of the low-mass cloud cores may be gravitationally bound and will collapse into stars, there are a lot of low mass cores in star-forming regions, and we can, therefore, expect that a considerable number of ELFs can form.

\section{Observational Properties of the First Cores}
We calculated the SEDs of first cores in the RHD simulations by performing post-processing radiation transfer. We solve the following radiation-transfer equation in one direction to draw an intensity distribution map for various inclination angles and wavelengths:
\begin{equation}
\frac{d I_\nu}{ds} = \rho \kappa_\nu(B_\nu(T) - I_\nu),
\end{equation}
where $B_\nu(T)$ is the Planck function of temperature $T$ and $\kappa_\nu$ is the monochromatic (absorption) opacity. Here, we assume that the temperature $T$ obtained in the RHD simulations is correct, and we ignore scattering, which is a valid assumption in mid-infrared or longer wavelengths. We adopt the monochromatic dust opacities provided by \citet{semenov}, using the homogeneous aggregates model of normal abundances. %We just use small values for gas opacities in high temperature region ($T\gtsim 1500{\rm K}$) where all the dust components evaporate, but this treatment causes no difference because such high temperature regions are too deeply embedded in the first core of very large optical depth.

\begin{figure}[p]
\scalebox{1}{\includegraphics{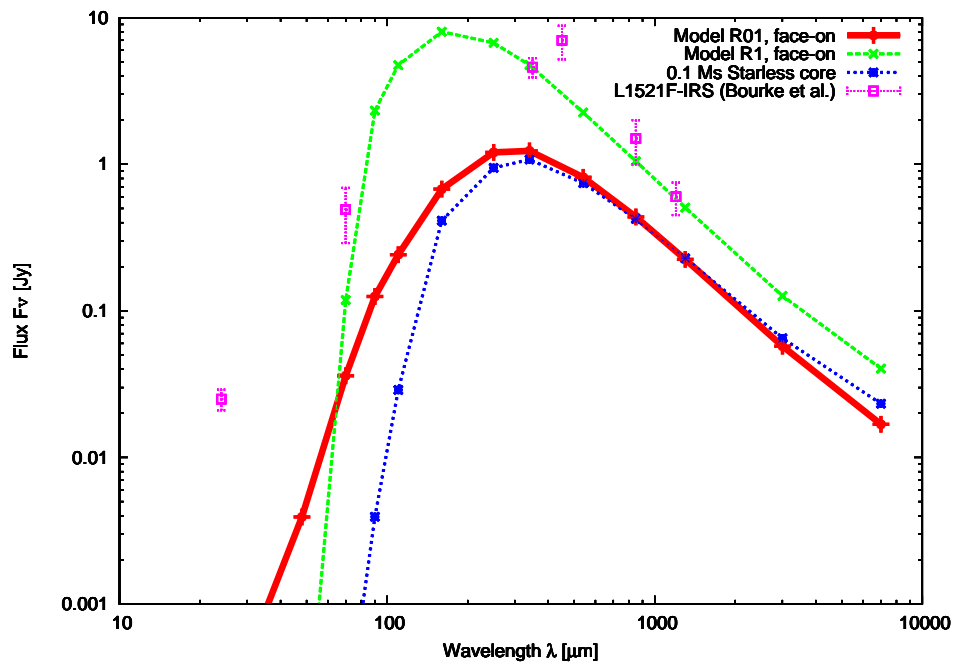}}
\caption{SEDs of the first core models at the same epoch as in Figure~\ref{fig1} in face-on configuration and a $0.1 M_\odot$ star-less core. The observed SED of L1521F-IRS \citep{bourke06} is also plotted. The distance towards the targets is $150 \  {\rm pc}$ and the aperture is $(1000 \  {\rm AU})^2$. First cores are more luminous than starless cores, especially in the far-infrared wavelengths. Model {\it R1} is brighter than Model {\it R01} in radio wavelengths. On the other hand, Model {\it R01} exceeds Model {\it R1} in the mid-infrared region.}
\label{sed}
\end{figure}

In Figure~\ref{sed}, we show the SEDs of our RHD models in the face-on configurations. For comparison, we also plot the SED of $0.1 M_\odot$ Bonnor-Ebert sphere, a model of a low mass starless core, and observed SED of L1521F-IRS \citep{bourke06}, which is a very low-luminosity object (VeLLO) in the Taurus molecular cloud. The distance towards the targets is set to $150 \  {\rm pc}$ and the SEDs are measured with a $(1000 \  {\rm AU})^2$ aperture. The flux in the far-infrared wavelengths increases when the first core forms and it can be observed with {\it Herschel}. Compared to L1521F-IRS,  Model {\it R1} has similar brightness in submillimeter region but it disagrees in mid-infrared region. In contrast, Model {\it R01} is fainter than L1521F-IRS in all the wavelengths. In radio wavelengths, in spite of having a mass 10 times larger than that of {\it R01}, {\it R1} is only about twice as bright as {\it R01}. However, the peak flux around the far-infrared wavelengths is significantly larger in {\it R1}. This is because it has a warmer first-core disk and envelope, as we mentioned before (Fig.~\ref{fig1}). In contrast, {\it R01} is more luminous in the mid-infrared wavelengths because it has a thinner envelope. If we follow the more long-term evolution of the low mass cloud model, this tendency is enhanced. Our models may explain the weak emission in mid-infrared wavelengths in some observed first core candidates \citep{en10}. We will discuss the time-evolution of observational properties of first cores in a subsequent paper.

\begin{figure}[p]
\scalebox{1}{\includegraphics{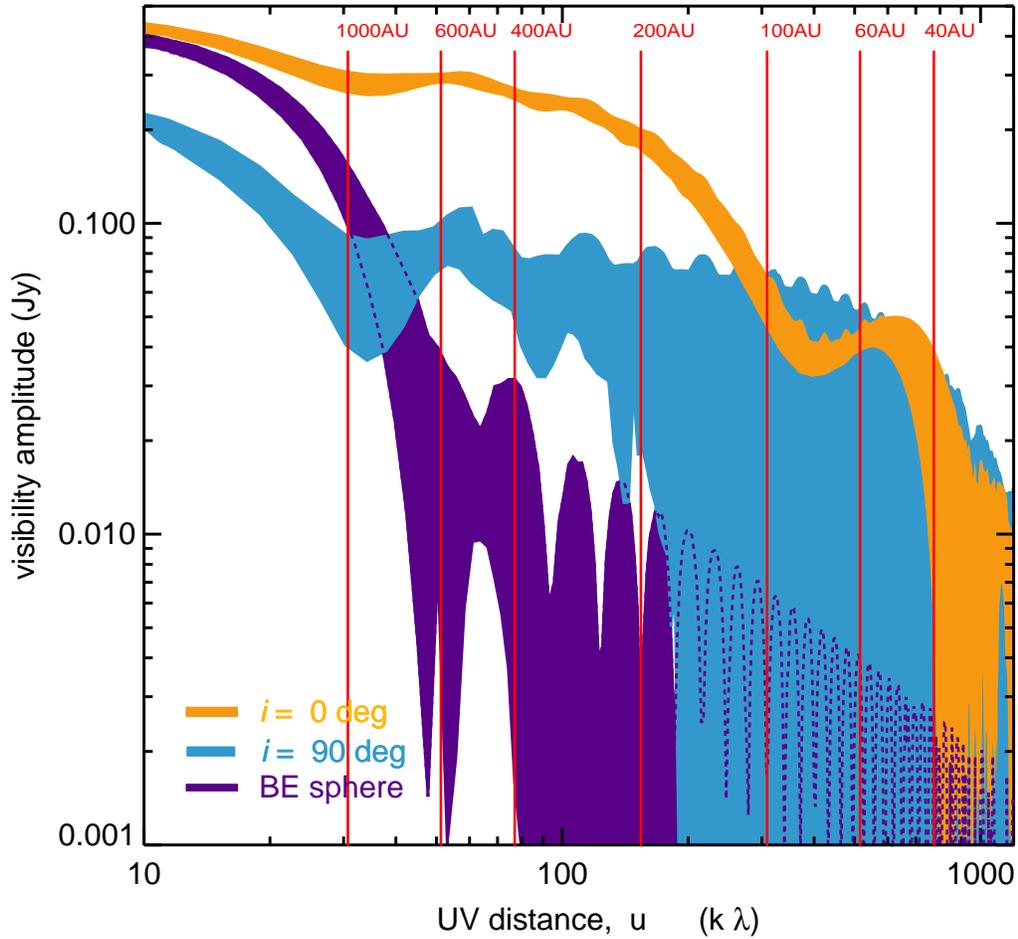}}
\caption{Visibility amplitude distributions of the $0.1 \  M_\odot$ first core model and a $0.1 \  M_\odot$ Bonnor-Ebert sphere as a model of a starless core. The first core model shows clearly shallower distribution compared to the starless core. The edge-on configuration shows more widely scattered visibility amplititude corresponding to its oblate morphology.}
\label{va}
\end{figure}

First cores emit larger flux in the infrared region than starless cores, but it is still difficult to identify the existence of a first core only from the SED, because it is so faint in the infrared wavelengths and the differences in radio wavelengths between the first cores and the starless cores are not so prominent. Another good method to distinguish first cores from starless cores is high resolution observation measuring visibility amplitude (Fourier components of the intensity map) distributions with (sub)millimeter interferometers. We show simulated visibility amplitude profiles of Model {\it R01} measured in $850 \  {\rm \mu m}$, obtained with the Common Astronomy Software Applications (CASA) in Figure~\ref{va}. The visibility amplitude of the starless core decreases steeply in the small scale because it contains no fine structure. In contrast, the first core clearly shows a shallow distribution. This feature is not solely seen in the ELF but common in first cores. This is firm evidence of the existence of the first cores. Current observations with SMA is not sufficient to resolve the first core directly, but sufficient to identify the first core. ALMA in full operation will reveal the detailed structure of the first core.

\section{Conclusions and Discussions}
We performed 3D RHD simulations of low-mass cloud cores and showed that first cores formed in very low mass ($0.1 M_\odot$) cloud cores live more than $10^4$ yrs. Those first cores have thin envelopes, therefore we name this first core an "Exposed Long-lifetime First core (ELF)." ELFs experience different evolution from ordinary first cores whose evolution is dominated by accretion from the natal core. ELFs use up the gas in the envelope in the early stages of its evolution, and then the mass and angular momentum redistribution in the disk control the evolution of ELFs. Radiation cooling plays a critical role there similar to the Kelvin-Helmholtz contraction, dominating the disk stability and angular momentum transport with the time-scale longer than the dynamical time-scale of accretion. Thus, the evolution of ELFs is different from that of the first cores formed in ordinary mass cloud cores, which have been well studied so far. The gas in ELFs does not behave adiabatically anymore and the barotropic approximation breaks down.

We also calculated the observational properties of ELFs. We found that in radio wavelengths ($300 \  {\rm \mu m}\sim$) they have slightly fainter but quite similar SEDs to ordinary mass first cores. 
This fact suggests that ELFs are detectable as well as ordinary first cores and they can be detected even with current instruments like SMA. On the other hand, ELFs are more luminous in mid-infrared region because they have less massive envelopes than those of typical first cores. 

Very low-mass cores will not collapse so often by self-gravity and the formation probability of ELFs may be low. However, ELFs are still worth considering because there are a number of low-mass cores both in observations \citep{motte,en07,rath09,gould} and in theoretical predictions \citep{pn02,hc08}. ELFs are long-lived compared to the dynamical time-scale of their natal cloud cores, so observation possibilities are higher than for usual first cores. Therefore, we can expect a considerable number of ELFs exist in the star-forming regions. Other mechanisms such as external pressure, cloud-cloud collision, radiation driven implosion \citep{mot07}, and dynamical ejection \citep{bate09} may help the formation of ELFs. We will study the effects of magnetic fields and realistic EOS in future works.

We thank Prof. Ryohei Kawabe and Prof. Nagayoshi Ohashi for their useful suggestions. Numerical computations were partly performed on NEC SX-9 at Center for Computational Astrophysics of National Astronomical Observatory of Japan, at Japan Aerospace Exploration Agency and at Osaka University. This work is partly supported by the Ministry of Education, Culture, Sports, Science and Technology (MEXT), Grants-in-Aid for Scientific Research, 21244021 (Tomisaka and MNM), 20540238 (TM), and 21740136 (MNM). K. Tomida is supported by the Research Fellowship from the Japan Society for the Promotion of Science (JSPS) for Young Scientists.

\clearpage

\end{document}